\documentclass[twocolumn,showpacs,preprintnumbers,amsmath,amssymb,superscriptaddress]{revtex4}
\usepackage[cp1251]{inputenc}
\usepackage[ukrainian]{babel}
\usepackage{bm}
\usepackage{epsfig,amssymb,amsmath,bm}
\usepackage[dvips]{color}

\begin{document}

\title{Factorization of the Lorentz transformations} 
\author{K. S. Karplyuk}
\email{karpks@hotmail.com}
 \affiliation{Department of Radiophysics, Taras Shevchenko University, Academic
Glushkov prospect 2, building 5, Kyiv 03122, Ukraine, karpks@hotmail.com}
\author{Myroslav I. Kozak}
\email{mko3ak@gmail.com}
 \affiliation{Department of Physics, Uzhhorod National University, 
Voloshyn st. 54, Ushhorod 88000, Ukraine}`
\author{O. O. Zhmudskyy}\email{ozhmudsky@physics.ucf.edu}
 \affiliation{Department of Physics, University of Central Florida, 4000 Central Florida Blvd. Orlando, FL, 32816, ozhmudsky@Knights.ucf.edu}
\begin{abstract}
The article shows how the factorization of an arbitrary Lorentz transformation is performed. That is, representation of an arbitrary 
Lorentz transformation as a sequence of spatial rotation and boost or boost and spatial rotation. Relations are obtained that determine 
the required boosts and turns.
\end{abstract}
\pacs{12., 12.20.-m, 12.15.-y, 13.66.-a}
\maketitle

\section{Introduction}
In quantum electrodynamics, the most convenient and natural form of Lorentz transformations is the hypercomplex form based on 16 
Dirac matrices \cite{bd}. In the hypercomplex representation, a scalar is associated with the matrix $a\hat{1}$, a pseudoscalar is associated with 
the matrix $a\hat{\iota}$, a 4-vector is associated with the matrix $a_\alpha \gamma^\alpha $, a 4-pseudovector is associated with the matrix 
$a_\alpha \pi^\alpha $, and an antisymmetric 
4-tensor of the second rank is associated with the matrix $a_{\alpha\beta}\sigma^{\alpha\beta}$. Here $\tilde{1}$ is the identity matrix $4\times 4$,
\begin{gather}
\gamma^\alpha\gamma^\beta+\gamma^\beta\gamma^\alpha=2g^{\alpha\beta},\\
\hat{\iota}=\gamma^0\gamma^1\gamma^2\gamma^3,\hspace{7mm}\pi^\alpha=\gamma^\alpha\hat{\iota},\hspace{7mm}
2\sigma^{\alpha\beta}=\gamma^\alpha\gamma^\beta-\gamma^\beta\gamma^\alpha.
\end{gather}  
As always, Greek indices take values $0$,$1$,$2$,$3$, Latin ones take values $1$,$2$,$3$.

Let's name the numbers
\begin{gather}
a\hat{1}+b\hat{\iota}+c_\alpha\gamma^\alpha+d_\alpha\pi^\alpha+\frac{1}{2}f_{\alpha\beta}\sigma^{\alpha\beta}
\end{gather}  
Dirac numbers \cite{cs}. The hypercomplex system of Dirac numbers contains a subsystem based on 8 matrices $\hat{1}$, $\hat{\iota}$, $\sigma^{\alpha\beta}$. 
The numbers of this subsystem have the form
\begin{gather}
a\hat{1}+b\hat{\iota}+\frac{1}{2}L_{\alpha\beta}\sigma^{\alpha\beta}=a\hat{1}+b\hat{\iota}+ \nonumber\\
+L_{01}\sigma^{01}+L_{02}\sigma^{02}+
L_{03}\sigma^{03}+L_{23}\sigma^{23}+L_{31}\sigma^{31}+L_{12}\sigma^{12}
\end{gather}

We will call these numbers the Lorentz numbers, since it is with their help that
Lorentz transformations are carried out. More precisely, with the help of matrix exponent
\begin{gather}
e^{\frac{1}{2}L_{\alpha\beta}\sigma^{\alpha\beta}}=\hat{1}+\frac{1}{2}L_{\alpha\beta}\sigma^{\alpha\beta}+
\frac{1}{2!}\frac{1}{2}L_{\alpha\beta}\sigma^{\alpha\beta}\frac{1}{2}L_{\alpha\beta}\sigma^{\alpha\beta}+\ldots=\nonumber\\=
\hat{1}+\frac{1}{2}L_{\alpha\beta}\sigma^{\alpha\beta}
-\frac{L^2}{2!}-\frac{L^2}{3!}\frac{1}{2}L_{\alpha\beta}\sigma^{\alpha\beta}+\frac{L^4}{4!}+\ldots=\nonumber\\=
\cos L+\frac{L_{\alpha\beta}\sigma^{\alpha\beta}}{2L}\sin L.
\end{gather}
Here we have used the equality  
\begin{gather}
\frac{1}{2}L_{\alpha\beta}\sigma^{\alpha\beta}\frac{1}{2}L_{\alpha\beta}\sigma^{\alpha\beta}=
-\frac{1}{2}(L^{\alpha\beta}L_{\alpha\beta}-\hat{\iota}L^{\alpha\beta}L_{\alpha\beta}^\diamond)=-L^2.
\end{gather}
and designation
\begin{gather}
L_{\alpha\beta}^\diamond=\frac{1}{2}\varepsilon_{\alpha\beta\mu\nu}L^{\mu\nu},\hspace{7mm}
\varepsilon^{0123}=1,\hspace{7mm}\varepsilon_{0123}=-1.
\end{gather}  
The tensor $L^\diamond_{\alpha\beta}$ dual to $L_{\alpha\beta}$ has components
\begin{gather}
L^\diamond_{01}=-L_{23},L^\diamond_{02}=-L_{31},L^\diamond_{03}=-L_{12},\nonumber\\ 
L^\diamond_{23}=L_{01},L^\diamond_{31}=L_{02},L^\diamond_{12}=L_{03}.
\end{gather}
The Lorentz transformation of scalars, pseudo-scalars, vectors, pseudo-vectors, and second-rank antisymmetric tensors is performed by the operations
\begin{gather}
a\hat{1}=e^{\frac{1}{2}L_{\alpha\beta}\sigma^{\alpha\beta}} a\hat{1} e^{-\frac{1}{2}L_{\alpha\beta}\sigma^{\alpha\beta}},\hspace{3mm}
b\hat{\iota}=e^{\frac{1}{2}L_{\alpha\beta}\sigma^{\alpha\beta}} b\hat{\iota} e^{-\frac{1}{2}L_{\alpha\beta}\sigma^{\alpha\beta}},\nonumber\\
c'_\alpha\gamma^\alpha=e^{\frac{1}{2}L_{\alpha\beta}\sigma^{\alpha\beta}} c_\alpha\gamma^\alpha e^{-\frac{1}{2}L_{\alpha\beta}\sigma^{\alpha\beta}},\nonumber\\
d'_\alpha\pi^\alpha=e^{\frac{1}{2}L_{\alpha\beta}\sigma^{\alpha\beta}} d_\alpha\pi^\alpha e^{-\frac{1}{2}L_{\alpha\beta}\sigma^{\alpha\beta}},\nonumber\\
f'_{\alpha\beta}\sigma^{\alpha\beta}=e^{\frac{1}{2}L_{\alpha\beta}\sigma^{\alpha\beta}} f_{\alpha\beta}\sigma^{\alpha\beta}
e^{-\frac{1}{2}L_{\alpha\beta}\sigma^{\alpha\beta}}.
\end{gather}
The Lorentz transformation of Dirac spinors is performed by the operation   
\begin{gather}
\psi'=e^{\frac{1}{2}L_{\alpha\beta}\sigma^{\alpha\beta}}\psi.
\end{gather}  
If the exponents in (9)-(10) have the form $\frac{1}{2}L_{kl}\sigma^{kl}$, then these formations are spatial rotations, but if they have the form 
$L_{0k}\sigma^{0k}$, then these transformations are boosts. In the general case, transformations are neither spatial rotations nor boosts. 
However, any Lorentz transformation can always be represented as a sequence of spatial rotation and boost or boost and spatial rotation. 
Below we obtain relations that allow us to do this in an arbitrary case.
\section{Biquaternion  representation of  the Lorentz transformations}
Hypercomplex Lorentz numbers are isomorphic with biquaternions. This makes it possible to use the well-known quaternion algebra to 
simplify manipulations with Lorentz transformations. To verify this isomorphism, we first note that
\begin{gather}
\sigma^{01}=-\hat{\iota}\sigma^{23}, \sigma^{02}=-\hat{\iota}\sigma^{31},\sigma^{03}=-\hat{\iota}\sigma^{12}.
\end{gather}
Therefore, (4) can be written as  
\begin{gather}
a+\hat{\iota}b+\frac{1}{2}L_{\alpha\beta}\sigma^{\alpha\beta}=
a+\hat{\iota}b-\hat{\iota}\frac{1}{2}L^\diamond_{kl}\sigma^{kl}+\frac{1}{2}L_{kl}\sigma^{kl}=\nonumber\\
=a+\hat{\iota}b+\frac{1}{2}(L_{kl}-\hat{\iota}L^\diamond_{kl})\sigma^{kl}=a+\hat{\iota}b+\frac{1}{2}l_{kl}\sigma^{kl}.
\end{gather}
Here 
\begin{gather}
l_{kl}=L_{kl}-\hat{\iota}L^\diamond_{kl}.
\end{gather}  
The algebra of matrices $\hat{1}$, $\sigma^{23}$, $\sigma^{31}$, $\sigma^{12}$ is isomorphic with the algebra of quaternions:
\[\mbox{\begin{tabular}{|c|c|c|c|} \hline \hline
 {$\times$}&$\sigma^{23}$&$\sigma^{31}$&$\sigma^{12}$\\
 \hline $\sigma^{23}$&-$\hat{1}$&$\sigma^{12}$&$-\sigma^{31}$\\
 \hline $\sigma^{31}$&$-\sigma^{12}$&-$\hat{1}$&$\sigma^{23}$\\ \hline
        $\sigma^{12}$&$\sigma^{31}$&$-\sigma^{23}$&-$\hat{1}$\\ \hline \hline
\end{tabular}}\hspace{13mm}\mbox{\begin{tabular}{|c|c|c|c|} \hline \hline
 {$\times$}&$\bm{i}$&$\bm{j}$&$\bm{k}$\\
 \hline $\bm{i}$&-1&$\bm{k}$&$-\bm{j}$\\
 \hline $\bm{j}$&$-\bm{k}$&-1&$\bm{i}$\\ \hline
        $\bm{k}$&$\bm{j}$&$-\bm{i}$&-1\\ \hline \hline
\end{tabular} }\]
Therefore the numbers 
\begin{gather}
a\hat{1}+\frac{1}{2}L_{kl}\sigma^{kl}
\end{gather}
can be thought of as quaternions, and the numbers   
\begin{gather}
a+\hat{\iota}b+\frac{1}{2}L_{\alpha\beta}\sigma^{\alpha\beta}=(a\hat{1}+\frac{1}{2}L_{kl}\sigma^{kl})+\hat{\iota}(b-\frac{1}{2}L_{kl}^\diamond\sigma^{kl})
\end{gather}  
like biquaternions \cite{bk}. That is, as a system of quaternions, expanded by introducing an additional unit $\hat{\iota}$. 

Mathematicians consider three possible options for introducing an additional unit: when $\hat{\iota}\cdot\hat{\iota}=-1$, 
when $\hat{\iota}\cdot\hat{\iota}=1$, and when $\hat{\iota}\cdot\hat{\iota}=0$. In the first case, the resulting numbers are 
called elliptic (ordinary) biquaternions, in the second, hyperbolic biquaternions, and in the third, parabolic ones. Since 
$\hat{\iota}\cdot\hat{\iota}=-1$, we are dealing with elliptic (ordinary) biquaternions.  

The numbers $a\hat{1}+b\hat{\iota}$ are isomorphic with complex numbers and commute with $\sigma^{\alpha\beta}$. 
We will call such numbers $\hat{\iota}$-complex numbers. Accordingly, biquaternions (15) can be considered as 
quaternions with $\hat{\iota}$-complex coefficients. In particular, exponent (5) can be written as   
\begin{gather}
e^{\frac{1}{2}L_{\alpha\beta}\sigma^{\alpha\beta}}=e^{\bm{l}\bm{\varsigma}}=1+\bm{l}\bm{\varsigma}+
\frac{1}{2!}\bm{l}\bm{\varsigma}\cdot\bm{l}\bm{\varsigma}+\frac{1}{3!}\bm{l}\bm{\varsigma}\cdot\bm{l}\bm{\varsigma}\cdot\bm{l}\bm{\varsigma}+\ldots=\nonumber\\
\cos l+\frac{\bm{l}\bm{\varsigma}}{l}\sin l.
\end{gather}
Here we use the notation  
\begin{gather}
\bm{l}\bm{\zeta}\equiv\frac{1}{2}L_{\alpha\beta}\sigma^{\alpha\beta}=\frac{1}{2}l_{kl}\sigma^{kl}=
\frac{1}{2}(L_{kl}-\hat{\iota}L^\diamond_{kl})\sigma^{kl}=\nonumber\\ 
(L_{23}-\hat{\iota}L^\diamond_{23})\sigma^{23}+(L_{31}-\hat{\iota}L^\diamond_{31})\sigma^{31}+(L_{12}-\hat{\iota}L^\diamond_{12})\sigma^{12}=\nonumber\\
(\bm{r}+\hat{\iota\bm{b}})\bm{\varsigma}= \\(r_x+\hat{\iota}b_x)\sigma^{23}+(r_y+\hat{\iota}b_y)\sigma^{31}+(r_z+\hat{\iota}b_z)\sigma^{12},\\
l^2=-\bm{l}\bm{\varsigma}\cdot\bm{l}\bm{\varsigma}=\bm{l}\cdot\bm{l}=(\bm{r}+\hat{\iota}\bm{b})^2=\bm{r}^2-\bm{b}^2+\hat{\iota}2\bm{b}\cdot\bm{r},\\
l=\sqrt{-\bm{l}\bm{\varsigma}\cdot\bm{l}\bm{\varsigma}}=\sqrt{\bm{l}\cdot\bm{l}}=\sqrt{\bm{r}^2-\bm{b}^2+\hat{\iota}2\bm{b}\cdot\bm{r}}.
\end{gather}
If $\bm{l}=\bm{r}$,  $\bm{b}=0$, then transformations (9)-(10) describe the space
direct rotation of the reference frame around the $\bm{r}$-axis by an angle $2r$. But if $\bm{l}=\hat{\iota}\bm{b}$,  $\bm{r}=0$, 
then transformations (9)-(10) describe the boost. Namely, if
$\frac{\bm{b}}{b}=\frac{\bm{v}}{v}$, $\tanh 2b=\frac{v}{c}$, then they will describe the transition to the reference frame that moves 
relative to the original system with a speed $\bm{v}$.  
In the general case, when $\bm{l}=\bm{r}+\hat{\iota}\bm{b}$, transformations (9) - (10) are neither spatial rotations nor boosts. 
However, as we will see below, they can always be represented as a sequence of spatial rotation and boost or boost and spatial rotation.  

Operations with biquaternion exponents (16) are much more convenient to perform if instead of $\hat{\iota}$-complex vectors $\bm{l}$,
 but $\hat{\iota}$-complex vectors $\bm{\lambda}$,  
 \begin{gather}
\bm{\lambda}(\bm{l})=\frac{\bm{l}}{l}\tan l.
\end{gather}
Obviously, in the case of a spatial rotation, when $\bm{l}$ is a $\hat{\iota}$-real vector $\bm{l}=\bm{r}$, the parameter $\bm{\lambda}$ 
is also a $\hat{\iota}$-real vector
\begin{gather}
\bm{\rho}(\bm{r})=\frac{\bm{r}}{r}\tan r.
\end{gather}  
In the boost case, when $\bm{l}$ is the $\hat{\iota}$-imaginary vector $\bm{l}=\hat{\iota}\bm{b}$, the parameter $\bm{\lambda}$ 
is also the $\hat{\iota}$-imaginary vector
\begin{gather}
\bm{\beta}(\hat{\iota}\bm{b})=\frac{\hat{\iota}\bm{b}}{b}\tanh b.
\end{gather}
If the parameter $\bm{\lambda}(\bm{l})$ is known, then the parameter $\bm{l}$ is determined by the relation
\begin{gather}
\bm{l}=\frac{\bm{\lambda}}{\sqrt{\bm{\lambda}\cdot\bm{\lambda}}}\arctan\sqrt{\bm{\lambda}\cdot\bm{\lambda}}=
\frac{\bm{\lambda}}{\lambda}\arctan\lambda, \nonumber\\ l=\arctan\sqrt{\bm{\lambda}\cdot\bm{\lambda}}=\arctan\lambda.
\end{gather}
When using the parameter $\bm{\lambda}(\bm{l})$, the exponent $e^{\bm{l}\bm{\varsigma}}$ takes the form
\begin{gather}
e^{\bm{l}\bm{\varsigma}}=\cos l+\frac{\bm{l}\bm{\varsigma}}{l}\sin l=\cos l(1+\frac{\bm{l}\bm{\varsigma}}{l}\tan l)= \nonumber\\
\frac{1}{\sqrt{1+\tan^2 l}}(1+\frac{\bm{l}\bm{\varsigma}}{l}\tan l)=\nonumber\\=
\frac{1+\bm{\lambda}\bm{\varsigma}}{\sqrt{1+\bm{\lambda}\cdot\bm{\lambda}}},
\end{gather}  
and the product of two exponents $e^{\bm{l}_2\bm{\varsigma}}e^{\bm{l}_1\bm{\varsigma}}$ is of the form
\begin{gather}
e^{\bm{l}_2\bm{\varsigma}}e^{\bm{l}_1\bm{\varsigma}}=\frac{(1+\bm{\lambda}_2\bm{\varsigma})(1+\bm{\lambda}_1\bm{\varsigma})}
{\sqrt{(1+\bm{\lambda}_2\cdot\bm{\lambda}_2)(1+\bm{\lambda}_1\cdot\bm{\lambda}_1)}}=\nonumber\\
=\frac{1-\bm{\lambda}_1\cdot\bm{\lambda}_2}
{\sqrt{(1+\bm{\lambda}_2\cdot\bm{\lambda}_2)(1+\bm{\lambda}_1\cdot\bm{\lambda}_1)}}
\Bigl(1+\frac{\bm{\lambda}_1+\bm{\lambda}_2+\bm{\lambda}_2\times\bm{\lambda}_1}{1-\bm{\lambda}_1\cdot\bm{\lambda}_2}\bm{\varsigma}\Bigr).
\end{gather}
Here $\bm{\lambda}_1=\bm{\lambda}(\bm{l}_1)$, $\bm{\lambda}_2=\bm{\lambda}(\bm{l}_2)$. Putting
\begin{gather}
\bm{\lambda}(\bm{l})=\frac{\bm{\lambda}_1+\bm{\lambda}_2+\bm{\lambda}_2\times\bm{\lambda}_1}{1-\bm{\lambda}_1\cdot\bm{\lambda}_2},
\end{gather}
we get  
\begin{gather}
1+\bm{\lambda}(\bm{l})\cdot\bm{\lambda}(\bm{l})=
\frac{(1+\bm{\lambda}_2\cdot\bm{\lambda}_2)(1+\bm{\lambda}_1\cdot\bm{\lambda}_1)}{(1-\bm{\lambda}_1\cdot\bm{\lambda}_2)^2}.
\end{gather}  
The product of the exponents takes the form of the exponent (16) with the exponent $\bm{l}\bm{\zeta}$ :
\begin{gather}
e^{\bm{l}_2\bm{\varsigma}}e^{\bm{l}_1\bm{\varsigma}}=\frac{1+\bm{\lambda}\bm{\varsigma}}{\sqrt{1+\bm{\lambda}\cdot\bm{\lambda}}}=e^{\bm{l}\bm{\varsigma}}.
\end{gather}
Relation (26) determines the rule for the composition of the parameters $\bm{\lambda}(\bm{l}_1)$ and $\bm{\lambda}(\bm{l}_2)$ when multiplying 
the exponents $e^{\bm{l}_2\bm{\varsigma}}e^{\bm{l}_1\bm{\varsigma}}$. Note that in \cite{f} the same relation was obtained for another parameter not related to biquaternions.
\section{Factorization of the Lorentz transformations} 
Any Lorentz transformation can be represented as a sequence of spatial rotation and boost or boost and spatial rotation. 
Accordingly, the exponent $e^{\bm{l}\bm{\varsigma}}$ can be represented as a product
\begin{gather}
e^{\bm{l}\bm{\varsigma}}=e^{\hat{\iota}\bm{b}\bm{\varsigma}}e^{\bm{r}\bm{\varsigma}},
\end{gather}
or 
\begin{gather}
e^{\bm{l}\bm{\varsigma}}=e^{\bm{\bm{r}'\bm{\varsigma}}}e^{\hat{\iota}\bm{b}'\bm{\varsigma}}.
\end{gather}  
Let's take a few steps to find the parameters of boosts $\bm{b}$, $\bm{b}'$ and rotations $\bm{r}$, $\bm{r}'$. 

At the first step, we multiply the left and right parts of equalities (29) -- (30) by the left and right parts of their Hermitian conjugate equalities
\begin{gather}
(e^{\bm{l}\bm{\varsigma}})^\dag=e^{-\bm{l}^\star\bm{\varsigma}}=e^{-\bm{r}\bm{\varsigma}}e^{\hat{\iota}\bm{b}\bm{\varsigma}}.
\end{gather}
or 
\begin{gather}
(e^{\bm{l}\bm{\varsigma}})^\dag=e^{-\bm{l}^\star\bm{\varsigma}}=e^{\hat{\iota}\bm{b}'\bm{\varsigma}}e^{-\bm{r}'\bm{\varsigma}}.
\end{gather}
Here ${}^\star$ denotes the  $\hat{\iota}$-complex conjugate: $\bm{l}=\bm{r}+\hat{\iota}\bm{b}$,  $\bm{l}^\star=\bm{r}-\hat{\iota}\bm{b}$.  

We multiply equality (29) by (31) from the right
\begin{gather}
e^{\bm{l}\bm{\varsigma}}e^{-\bm{l}^\star\bm{\varsigma}}=
e^{\hat{\iota}\bm{b}\bm{\varsigma}}e^{\bm{r}\bm{\varsigma}}
e^{-\bm{r}\bm{\varsigma}}e^{\hat{\iota}\bm{b}\bm{\varsigma}}=e^{\hat{\iota}2\bm{b}\bm{\varsigma}},
\end{gather}
and equality (30) is multiplied by (32) from the left 
\begin{gather}
e^{-\bm{l}^\star\bm{\varsigma}}e^{\bm{l}\bm{\varsigma}}=
e^{\hat{\iota}\bm{b}'\bm{\varsigma}}e^{\bm{\bm{r}'\bm{\varsigma}}}e^{-\bm{\bm{r}'\bm{\varsigma}}}e^{\hat{\iota}\bm{b}'\bm{\varsigma}}=
e^{\hat{\iota}\bm{b}'\bm{\varsigma}}e^{\hat{\iota}\bm{b}'\bm{\varsigma}}=e^{\hat{\iota}2\bm{b}'\bm{\varsigma}}.
\end{gather} 
Using (26), we find the parameters $\bm{\lambda}(\hat{\iota}2\bm{b})$ and $\bm{\lambda}(\hat{\iota}2\bm{b}')$ corresponding to the 
parameters $\bm{l}=\hat{\iota}2\bm{b}$ and $\bm{l}'=\hat{\iota}2\bm{b}'$:
\begin{gather}
\bm{\lambda}(\hat{\iota}2\bm{b})=\frac{\bm{\lambda}(\bm{l})+\bm{\lambda}^\star(-\bm{l})+\bm{\lambda}(\bm{l})\times\bm{\lambda}^\star(-\bm{l})}
{1-\bm{\lambda}(\bm{l})\cdot\bm{\lambda}^\star(-\bm{l})}= \nonumber\\
\frac{\bm{\lambda}(\bm{l})-\bm{\lambda}^\star(\bm{l})-\bm{\lambda}(\bm{l})\times\bm{\lambda}^\star(\bm{l})}
{1+\bm{\lambda}(\bm{l})\cdot\bm{\lambda}^\star(\bm{l})},\\
\bm{\lambda}(\hat{\iota}2\bm{b}')=\frac{\bm{\lambda}(\bm{l})+\bm{\lambda}^\star(-\bm{l})+\bm{\lambda}^\star(-\bm{l})\times\bm{\lambda}(\bm{l})}
{1-\bm{\lambda}(\bm{l})\cdot\bm{\lambda}^\star(-\bm{l})}= \nonumber\\
\frac{\bm{\lambda}(\bm{l})-\bm{\lambda}^\star(\bm{l})-\bm{\lambda}^\star(\bm{l})\times\bm{\lambda}(\bm{l})}
{1+\bm{\lambda}(\bm{l})\cdot\bm{\lambda}^\star(\bm{l})}.
\end{gather}
At the second step, we will find the parameters $\bm{\lambda}(\hat{\iota}\bm{b})$ and  $\bm{\lambda}(\hat{\iota}\bm{b}')$ we need. 
They differ from $\bm{\lambda}(\hat{\iota}2\bm{b})$ and $\bm{\lambda}(\hat{\iota}2\bm{b}')$ by factors
\begin{gather}
\frac{\tan\sqrt{\hat{\iota}\bm{b}\cdot\hat{\iota}\bm{b}}}{\tan\sqrt{\hat{\iota} 2\bm{b}\cdot\hat{\iota} 2\bm{b}}}\hspace{7mm} \mbox{and}
\hspace{7mm}\frac{\tan\sqrt{\hat{\iota}\bm{b}'\cdot\hat{\iota}\bm{b}'}}{\tan\sqrt{\hat{\iota} 2\bm{b}'\cdot\hat{\iota} 2\bm{b}'}}.
\end{gather}
To find these factors, we use the trigonometric  equality 
\begin{gather}
\frac{\tan z}{\tan 2z}=\frac{1}{1+\sqrt{1+\tan^2 2z}}.
\end{gather}
Thus  
\begin{gather}
\frac{\tan\sqrt{\hat{\iota}\bm{b}\cdot\hat{\iota}\bm{b}}}{\tan\sqrt{\hat{\iota} 2\bm{b}\cdot\hat{\iota} 2\bm{b}}}=
\frac{1}{1+\sqrt{1+\tan^2\sqrt{\hat{\iota}2\bm{b}\cdot\hat{\iota}2\bm{b}}}}= \nonumber\\
\frac{1}{1+\sqrt{1+\bm{\lambda}(\hat{\iota}2\bm{b})\cdot\bm{\lambda}(\hat{\iota}2\bm{b})}},\\
\frac{\tan\sqrt{\hat{\iota}\bm{b}'\cdot\hat{\iota}\bm{b}'}}{\tan\sqrt{\hat{\iota} 2\bm{b}'\cdot\hat{\iota} 2\bm{b}'}}=\!
\frac{1}{1\!+\sqrt{1\!+\tan^2\sqrt{\hat{\iota}2\bm{b}'\cdot\hat{\iota}2\bm{b}'}}}= \nonumber\\
\frac{1}{1\!+\sqrt{1\!+\bm{\lambda}(\hat{\iota}2\bm{b}')\cdot\bm{\lambda}(\hat{\iota}2\bm{b}')}}.
\end{gather}
Let's calculate  
\begin{gather}
\bm{\lambda}(\hat{\iota}2\bm{b})\cdot\bm{\lambda}(\hat{\iota} 2\bm{b})=
\bm{\lambda}(\hat{\iota}2\bm{b}')\cdot\bm{\lambda}(\hat{\iota} 2\bm{b}')=\nonumber\\=
\frac{\bm{\lambda}^2(\bm{l})+\bm{\lambda}^{\star 2}(\bm{l})-
2\bm{\lambda}(\bm{l})\cdot\bm{\lambda}^{\star}(\bm{l})+\bm{\lambda}^2(\bm{l})\bm{\lambda}^{\star 2}(\bm{l})-
[\bm{\lambda}(\bm{l})\cdot\bm{\lambda}^\star(\bm{l})]^2}
{[1+\bm{\lambda}(\bm{l})\cdot\bm{\lambda}^\star(\bm{l})]^2}=\nonumber\\=\!
\frac{[1+\bm{\lambda}^2(\bm{l})][1+\bm{\lambda}^{\star 2}(\bm{l})]-[1+\bm{\lambda}(\bm{l})\cdot\bm{\lambda}^\star(\bm{l})]^2}
{[1+\bm{\lambda}(\bm{l})\cdot\bm{\lambda}^\star(\bm{l})]^2}= \nonumber\\
\frac{[1+\bm{\lambda}^2(\bm{l})][1+\bm{\lambda}^{\star 2}(\bm{l})]}
{[1+\bm{\lambda}(\bm{l})\cdot\bm{\lambda}^\star(\bm{l})]^2}-1.
\end{gather}
Respectively  
\begin{gather}
\frac{\tan\sqrt{\hat{\iota}\bm{b}\cdot\hat{\iota}\bm{b}}}{\tan\sqrt{\hat{\iota} 2\bm{b}\cdot\hat{\iota} 2\bm{b}}}=
\frac{\tan\sqrt{\hat{\iota}\bm{b}'\cdot\hat{\iota}\bm{b}'}}{\tan\sqrt{\hat{\iota} 2\bm{b}'\cdot\hat{\iota} 2\bm{b}'}}=\nonumber\\=
\frac{1+\bm{\lambda}(\bm{l})\cdot\bm{\lambda}^\star(\bm{l})}{1+\bm{\lambda}(\bm{l})\cdot\bm{\lambda}^\star(\bm{l})+
\sqrt{[1+\bm{\lambda}^2(\bm{l})][1+\bm{\lambda}^2(\bm{l})]^\star}}.
\end{gather}  
Thus, the parameters $\bm{\lambda}(\hat{\iota}\bm{b})$ and $\bm{\lambda}(\hat{\iota}\bm{b}')$ corresponding to the parameters 
$\bm{l}=\hat{\iota}\bm{b}$ and  $\bm{l}'=\hat{\iota}\bm{b}'$, have the form
\begin{gather}
\bm{\lambda}(\hat{\iota}\bm{b})= 
\bm{\lambda}(\hat{\iota}2\bm{b})\frac{\tan\sqrt{\hat{\iota}\bm{b}\cdot\hat{\iota}\bm{b}}}{\tan\sqrt{\hat{\iota} 2\bm{b}\cdot\hat{\iota} 2\bm{b}}}=
\nonumber\\=
\frac{\bm{\lambda}(\bm{l})-\bm{\lambda}^\star(\bm{l})-\bm{\lambda}(\bm{l})\times\bm{\lambda}^\star(\bm{l})}
{1+\bm{\lambda}(\bm{l})\!\cdot\!\bm{\lambda}^\star(\bm{l})+\sqrt{[1+\bm{\lambda}^2(\bm{l})][1+\bm{\lambda}^2(\bm{l})]^\star}},\\
\bm{\lambda}(\hat{\iota}\bm{b}')=\!
\bm{\lambda}(\hat{\iota}2\bm{b}')\frac{\tan\sqrt{\hat{\iota}\bm{b}'\!\cdot\!\hat{\iota}\bm{b}'}}{\tan\sqrt{\hat{\iota} 2\bm{b}'\!\cdot\!\hat{\iota} 2\bm{b}'}}\!=\!
\nonumber\\=
\frac{\bm{\lambda}(\bm{l})-\bm{\lambda}^\star(\bm{l})+\bm{\lambda}(\bm{l})\times\bm{\lambda}^\star(\bm{l})}
{1\!+\!\bm{\lambda}(\bm{l})\!\cdot\!\bm{\lambda}^\star(\bm{l})\!+\!\sqrt{[1+\bm{\lambda}^2(\bm{l})][1+\bm{\lambda}^2(\bm{l})]^\star}}.
\end{gather}  
As it should be, the parameters $\bm{\lambda}(\hat{\iota}\bm{b})$ and $\bm{\lambda}(\hat{\iota}\bm{b}')$ are $\hat{\iota}$-imaginary vectors. 
The direction of these vectors depends on which operation - turn or boost - is performed first, and which is second. 
The magnitude of the vectors $\bm{\lambda}(\hat{\iota}\bm{b})$ and $\bm{\lambda}(\hat{\iota}\bm{b}')$ does not depend on this.  

At the third step, we find the exponents describing spatial rotations. To do this, we multiply (29) 
by $e^{-\hat{\iota}\bm{b}\bm{\varsigma}}$ on the left side, and (30) by $e^{-\hat{\iota}\bm{b}'\bm{\varsigma}}$ on the right side:
\begin{gather}
e^{-\hat{\iota}\bm{b}\bm{\varsigma}}e^{\bm{l}\bm{\varsigma}}=
e^{-\hat{\iota}\bm{b}\bm{\varsigma}}e^{\hat{\iota}\bm{b}\bm{\varsigma}}e^{\bm{r}\bm{\varsigma}}=e^{\bm{r}\bm{\varsigma}},\\
e^{\bm{l}\bm{\varsigma}}e^{-\hat{\iota}\bm{b}'\bm{\varsigma}}=
e^{\bm{\bm{r}'\bm{\varsigma}}}e^{\hat{\iota}\bm{b}'\bm{\varsigma}}e^{-\hat{\iota}\bm{b}'\bm{\varsigma}}=e^{\bm{\bm{r}'\bm{\varsigma}}}.
\end{gather}
We use (26) again and after long but not complicated transformations we find the parameters $\bm{\lambda}(\bm{r})$ and  
$\bm{\lambda}(\bm{r}')$ corresponding to the parameters $\bm{r}$ and $\bm{r}'$:
\begin{widetext}
\begin{gather}
\bm{\lambda}(\bm{r})=\frac{\bm{\lambda}(\bm{l})+\bm{\lambda}(-\hat{\iota}\bm{b})+\bm{\lambda}(-\hat{\iota}\bm{b})\times\bm{\lambda}(\bm{l})}
{1-\bm{\lambda}(\bm{l})\cdot\bm{\lambda}(-\hat{\iota}\bm{b})}=
\frac{\bm{\lambda}(\bm{l})-\bm{\lambda}(\hat{\iota}\bm{b})+\bm{\lambda}(\bm{l})\times\bm{\lambda}(\hat{\iota}\bm{b})}
{1+\bm{\lambda}(\bm{l})\cdot\bm{\lambda}(\hat{\iota}\bm{b})}=\nonumber\\
=\!\Bigl\{\bm{\lambda}(\bm{l})+\bm{\lambda}(\bm{l})[\bm{\lambda}(\bm{l})\cdot\!\bm{\lambda}^\star(\bm{l})]+
\bm{\lambda}(\bm{l})\sqrt{[1+\bm{\lambda}^2(\bm{l})][1+\bm{\lambda}^2(\bm{l})]^\star}-\!\bm{\lambda}(\bm{l})+\bm{\lambda}^\star(\bm{l})+\nonumber\\+
\bm{\lambda}(\bm{l})\times\bm{\lambda}^\star(\bm{l})-\bm{\lambda}(\bm{l})\times\bm{\lambda}^\star(\bm{l})
-\bm{\lambda}(\bm{l})[\bm{\lambda}(\bm{l})\!\cdot\!\bm{\lambda}^\star(\bm{l})]+
\bm{\lambda}^\star(\bm{l})[\bm{\lambda}(\bm{l})\!\cdot\!\bm{\lambda}(\bm{l})]\Bigr\}\times\nonumber\\
\times\Bigl\{1+\bm{\lambda}(\bm{l})\!\cdot\!\bm{\lambda}^\star(\bm{l})+\sqrt{[1+\bm{\lambda}^2(\bm{l})][1+\bm{\lambda}^2(\bm{l})]^\star}\Bigr\}^{-1}
\{1+\bm{\lambda}(\bm{l})\cdot\bm{\lambda}(\hat{\iota}\bm{b})\}^{-1}=\nonumber\\
=\frac{\bm{\lambda}(\bm{l})\sqrt{[1+\bm{\lambda}^2(\bm{l})][1+\bm{\lambda}^2(\bm{l})]^\star}+
\bm{\lambda}^\star(\bm{l})[1+\bm{\lambda}(\bm{l})\!\cdot\!\bm{\lambda}(\bm{l})]}
{1+\bm{\lambda}(\bm{l})\!\cdot\!\bm{\lambda}^\star(\bm{l})+\sqrt{[1+\bm{\lambda}^2(\bm{l})][1+\bm{\lambda}^2(\bm{l})]^\star}}\times\nonumber\\
\times\frac{1+\bm{\lambda}(\bm{l})\!\cdot\!\bm{\lambda}^\star(\bm{l})+\sqrt{[1+\bm{\lambda}^2(\bm{l})][1+\bm{\lambda}^2(\bm{l})]^\star}}
{1+\bm{\lambda}(\bm{l})\!\cdot\!\bm{\lambda}^\star(\bm{l})+\sqrt{[1+\bm{\lambda}^2(\bm{l})][1+\bm{\lambda}^2(\bm{l})]^\star}
+\bm{\lambda}(\bm{l})\!\cdot\!\bm{\lambda}(\bm{l})-\bm{\lambda}(\bm{l})\!\cdot\!\bm{\lambda}^\star(\bm{l})}=\nonumber\\
=\frac{\bm{\lambda}(\bm{l})\sqrt{[1+\bm{\lambda}^2(\bm{l})]^\star}+\bm{\lambda}^\star(\bm{l})\sqrt{[1+\bm{\lambda}^2(\bm{l})]}}
{\sqrt{[1+\bm{\lambda}^2(\bm{l})]^\star}+\sqrt{[1+\bm{\lambda}^2(\bm{l})]}},
\end{gather}
\begin{gather}
\bm{\lambda}(\bm{r}')=\frac{\bm{\lambda}(\bm{l})+\bm{\lambda}(-\hat{\iota}\bm{b}')+\bm{\lambda}(\bm{l})\!\times\!\bm{\lambda}(-\hat{\iota}\bm{b}')}
{1-\bm{\lambda}(\bm{l})\cdot\bm{\lambda}(-\hat{\iota}\bm{b}')}=
\frac{\bm{\lambda}(\bm{l})\!-\!\bm{\lambda}(\hat{\iota}\bm{b}')\!-\!\bm{\lambda}(\bm{l})\!\times\!\bm{\lambda}(\hat{\iota}\bm{b}')}
{1+\bm{\lambda}(\bm{l})\cdot\bm{\lambda}(\hat{\iota}\bm{b}')}=\nonumber\\=
\Bigl\{\bm{\lambda}(\bm{l})+\bm{\lambda}(\bm{l})[\bm{\lambda}(\bm{l})\cdot\!\bm{\lambda}^\star(\bm{l})]\!+\!
\bm{\lambda}(\bm{l})\sqrt{[1+\bm{\lambda}^2(\bm{l})][1+\bm{\lambda}^2(\bm{l})]^\star}-\!\bm{\lambda}(\bm{l})+\bm{\lambda}^\star(\bm{l})-\nonumber\\-
\bm{\lambda}(\bm{l})\times\bm{\lambda}^\star(\bm{l})+\bm{\lambda}(\bm{l})\times\bm{\lambda}^\star(\bm{l})
-\bm{\lambda}(\bm{l})[\bm{\lambda}(\bm{l})\!\cdot\!\bm{\lambda}^\star(\bm{l})]+
\bm{\lambda}^\star(\bm{l})[\bm{\lambda}(\bm{l})\!\cdot\!\bm{\lambda}(\bm{l})]\Bigr\}\times\nonumber\\
\times\Bigl\{1+\bm{\lambda}(\bm{l})\!\cdot\!\bm{\lambda}^\star(\bm{l})+\sqrt{[1+\bm{\lambda}^2(\bm{l})][1+\bm{\lambda}^2(\bm{l})]^\star}\Bigr\}^{-1}
\{1+\bm{\lambda}(\bm{l})\cdot\bm{\lambda}(\hat{\iota}\bm{b}')\}^{-1}=\nonumber\\
=\frac{\bm{\lambda}(\bm{l})\sqrt{[1+\bm{\lambda}^2(\bm{l})][1+\bm{\lambda}^2(\bm{l})]^\star}+
\bm{\lambda}^\star(\bm{l})[1+\bm{\lambda}(\bm{l})\!\cdot\!\bm{\lambda}(\bm{l})]}
{1+\bm{\lambda}(\bm{l})\!\cdot\!\bm{\lambda}^\star(\bm{l})+\sqrt{[1+\bm{\lambda}^2(\bm{l})][1+\bm{\lambda}^2(\bm{l})]^\star}}\times\nonumber\\
\times\frac{1+\bm{\lambda}(\bm{l})\!\cdot\!\bm{\lambda}^\star(\bm{l})+\sqrt{[1+\bm{\lambda}^2(\bm{l})][1+\bm{\lambda}^2(\bm{l})]^\star}}
{1+\bm{\lambda}(\bm{l})\!\cdot\!\bm{\lambda}^\star(\bm{l})+\sqrt{[1+\bm{\lambda}^2(\bm{l})][1+\bm{\lambda}^2(\bm{l})]^\star}
+\bm{\lambda}(\bm{l})\!\cdot\!\bm{\lambda}(\bm{l})-\bm{\lambda}(\bm{l})\!\cdot\!\bm{\lambda}^\star(\bm{l})}=\nonumber\\
=\frac{\bm{\lambda}(\bm{l})\sqrt{[1+\bm{\lambda}^2(\bm{l})]^\star}+\bm{\lambda}^\star(\bm{l})\sqrt{[1+\bm{\lambda}^2(\bm{l})]}}
{\sqrt{[1+\bm{\lambda}^2(\bm{l})]^\star}+\sqrt{[1+\bm{\lambda}^2(\bm{l})]}}.
\end{gather}
\end{widetext}
As expected, the parameters $\bm{\lambda}(\bm{r})$ and $\bm{\lambda}(\bm{r}')$ are $\hat{\iota}$-real vectors. Neither the magnitude nor 
the direction of these vectors depends on the sequence in which the turn and boost are performed. 

Having obtained the parameters $\bm{\lambda}(\hat{\iota}\bm{b})$, $\bm{\lambda}(\hat{\iota}\bm{b}')$, $\bm{\lambda}(\bm{r})$ and 
$\bm{\lambda}(\bm{r}')$, we can use relations (23) to pass to the parameters $\bm{l}$, $\bm{l}'$, $\bm{r}$,  $\bm{r}'$ and represent an 
arbitrary transformation Lorentz in the form (29) or (30). It is even simpler to express the exponents in (29) -- (30) directly in terms of 
$\bm{\lambda}(\hat{\iota}\bm{b})$, $\bm{\lambda}(\hat{\iota}\bm{b}')$, $\bm{\lambda}(\bm{r})$ and $\bm{\lambda}(\bm{r}')$ using relation (24).  
\section{Summary}
The article shows how any Lorentz transformation can be represented as a sequence of spatial rotation and boost or boost and spatial rotation 
transformations. Relations are found that determine the parameters of such turns and boosts. Representing an arbitrary Lorentz transformation
 in the form of rotation and boost or boost and rotation makes it possible to give a physical meaning to this transformation and to analyze it.  
 \section{Acknowledgements}
The authors would like to thank  Prof. Evgeniy Tolkachev for stimulating discussions.


\end{document}